\documentclass[preprint,superscriptaddress,showpacs,preprintnumbers,amsmath,amssymb]{revtex4}

\usepackage{graphicx}
\usepackage{dcolumn}
\usepackage{bm}

\def\lesssim{\ \raise.3ex\hbox{$<$}\kern-0.8em\lower.7ex\hbox{$\sim$}\ }
\def\gesim{\ \raise.3ex\hbox{$>$}\kern-0.8em\lower.7ex\hbox{$\sim$}\ }

\begin{document}

\title{Single-particle properties and pseudogap effects in the BCS-BEC
crossover regime of an ultracold Fermi gas above $T_{\rm c}$}

\author{Shunji Tsuchiya}%
\affiliation{Department of Physics, Keio University, 3-14-1 Hiyoshi, Kohoku-ku, Yokohama 223-8522, Japan}
\affiliation{CREST(JST), 4-1-8 Honcho, Saitama 332-0012, Japan}

\author{Ryota Watanabe}%
\affiliation{Department of Physics, Keio University, 3-14-1 Hiyoshi, Kohoku-ku, Yokohama 223-8522, Japan}

\author{Yoji Ohashi}%
\affiliation{Department of Physics, Keio University, 3-14-1 Hiyoshi, Kohoku-ku, Yokohama 223-8522, Japan}
\affiliation{CREST(JST), 4-1-8 Honcho, Saitama 332-0012, Japan}

\date{\today}

\begin{abstract}

We investigate strong-coupling effects on normal state properties of an
 ultracold Fermi gas. Within the framework of $T$-matrix approximation
 in terms of pairing fluctuations, we calculate the single-particle
 density of states (DOS), as well as the spectral weight, over the
 entire BCS-BEC crossover region above the superfluid phase transition
 temperature $T_{\rm c}$. Starting from the weak-coupling BCS regime, we
 show that the so-called pseudogap develops in DOS above $T_{\rm c}$,
 which becomes remarkable in the crossover region. The pseudogap
 structure continuously changes into a fully gapped one in the
 strong-coupling BEC regime, where the gap energy is directly related to
 the binding energy of tightly bound molecules. We determine the
 pseudogap temperature $T^*$ where the dip structure in DOS
 vanishes. The value of $T^*$ is shown to be very different from another
 characteristic temperature $T^{**}$ where a BCS-type double peak
 structure disappears in the spectral weight. While one finds
 $T^*>T^{**}$ in the BCS regime, $T^{**}$ becomes higher than $T^*$ in
 the crossover region and BEC regime. 
 Including this, we determine the pseudogap region in the phase diagram
 of ultracold Fermi gases. Our results would be useful in the search for
 the pseudogap region in ultracold $^6$Li and $^{40}$K Fermi gases. 
\end{abstract}

\maketitle

\section{introduction}
Ultracold atomic Fermi gases provide unique opportunities to investigate
the crossover from the Bardeen-Cooper-Shrieffer (BCS) type
superfluids to the Bose-Einstein condensation (BEC) of tightly bound
molecules\cite{Eagles,Leggett,Nozieres,SadeMelo} in a unified
manner\cite{Regal,Zwierlein,Bartenstein,Kinast,Ketterle}.
One of the key ingredients to achieve this BCS-BEC crossover
in Fermi gases is a Feshbach resonance\cite{Ketterle}, which allows one
to tune the pairing interaction from the weak-coupling BCS limit to the
strong coupling BEC limit\cite{Timmermans,Holland,Ohashi,Bloch,Giorgini}.
Since the BCS-BEC crossover is a fundamental many-body problem,
it has recently attracted much
attention, not only in cold atom physics, but also in various research
fields, such as condensed matter physics and high energy physics.
In particular, this system is expected to be helpful for further
understanding of high-$T_{\rm c}$ cuprates, which has been one of the
most challenging problems in condensed matter physics\cite{Lee}.
\par
In the under-doped regime of high-$T_{\rm c}$ cuprates, the so-called
pseudogap phenomenon has been extensively studied\cite{Lee,Fischer}. In
this phenomenon, the single-particle density of states (DOS) in the
normal state exhibits a dip structure around the Fermi energy. 
The temperature at which the pseudogap appears is referred to as the
pseudogap temperature $T^*$, which is higher than the superconducting phase transition temperature
$T_{\rm c}$. In the region between $T^*$ and $T_{\rm c}$, various
anomalies have been observed in physical quantities, such as nuclear
spin-lattice relaxation rate (NMR-$T_1^{-1}$)\cite{Yasuoka}, and
angle-resolved photoemission spectroscopy (ARPES)\cite{Damascelli}. 
As the origin of the pseudogap,
possibility of preformed pairs due to strong pairing fluctuations has been
proposed\cite{Randeria,Singer,Janko,Rohe,Yanase,Perali}. 
However, because of the complexity of high-$T_{\rm c}$ cuprates, other
scenarios have been also discussed, such as antiferromagnetic spin
fluctuations\cite{Pines,Kampf} and a hidden order\cite{Chakravarty}. 
Thus, a simple system only having strong pairing fluctuations would be
helpful to confirm whether or not preformed pairs are responsible
for the pseudogap formation in high-$T_{\rm c}$ cuprates.
\par
In this regard, the cold Fermi gas system meets this demand. This
system is much cleaner and simpler than high-$T_{\rm c}$ cuprates, and
the pairing mechanism associated with a Feshbach resonance has been well
understood. The BCS-BEC crossover is dominated by strong pairing
fluctuations, so that one can focus on how they affect physical quantities. 
Indeed, effects of pairing fluctuations on single-particle spectral
weight have been theoretically studied by many
researchers\cite{Rohe,Yanase,Janko,Perali,Pieri,Bruun,Massignan,Chen,Haussmann}.
They clarified that pairing fluctuations lead to a BCS-type double peak
structure in the spectral weight above $T_{\rm c}$, which is a signature
of pseudogap phenomenon. 
They also found that the two peaks in the spectral weight merge into a
single peak at high temperatures. In Ref.~\cite{Perali}, detailed
analysis on the spectral weight above $T_{\rm c}$ has been carried out
over the entire BCS-BEC crossover, and, in the BEC regime, the deviation
from the BCS-type behaviors due to an asymmetric double peak structure
has been pointed out. 
Since a photoemission-type experiment has recently become
possible in cold atom physics\cite{Stewart}, we can now examine
strong-coupling effects on single-particle excitations within the
current experimental technology.
Although cold Fermi gases are not exactly the same as
high-$T_{\rm c}$ cuprates (e.g., pairing symmetry), the study of
pseudogap phenomenon in cold Fermi gases is expected to be useful for
further understanding of the underdoped regime of high-$T_{\rm c}$
cuprates.
\par
In this paper, we investigate pseudogap behaviors of an ultracold Fermi
gas above $T_{\rm c}$. Including pairing fluctuations within the $T$-matrix
approximation developed in Refs.~\cite{Rohe,Perali}, we systematically
examine how the pseudogap develops in DOS, as well as the spectral
weight, over the entire BCS-BEC crossover region. We determine the
pseudogap temperature $T^*$ at which the dip structure in DOS vanishes. We
show that $T^*$ is quite different from the temperature $T^{**}$ where
the double peak structure in the spectral weight disappears. In 
the BCS regime, we find that $T^*>T^{**}$. However, $T^{**}$ becomes
higher than $T^*$ in the crossover region and BEC regime. 
Including this, we determine the pseudogap region in the BCS-BEC
crossover phase diagram in terms of temperature and the strength of pairing interaction.
\par
This paper is organized as follows. In Sec.~\ref{Model}, we explain our model and
formulation to study pseudogap in DOS and spectral weight. In
Sec.~\ref{results}, we examine the pseudogap structure in DOS. Here, we show how
the pseudogapped DOS continuously changes into fully gapped one, as
one passes through the BCS-BEC crossover region. We determine
the pseudogap temperature $T^*$ from the temperature dependence of
DOS. In Sec.~\ref{spweight}, we examine strong-coupling effects on the spectral
weight. We introduce another pseudogap temperature $T^{**}$ from the
temperature dependence of spectral weight. We also discuss difference
between $T^*$ and $T^{**}$. Throughout this paper, we take $\hbar=k_{\rm
B}=1$.

\section{Model and formalism}
\label{Model}

We consider a three-dimensional uniform Fermi gas, consisting of two
atomic hyperfine states described by pseudospin
$\sigma=\uparrow,\downarrow$. So far, all the experiments on cold Fermi
gases are using a broad Feshbach resonance to tune the strength of a
pairing
interaction\cite{Regal,Zwierlein,Bartenstein,Kinast,Ketterle}. In this
case, detailed Feshbach-induced pairing mechanism is known to be
not crucial as far as we consider the interesting BCS-BEC crossover
regime, and one can safely use the ordinary single-channel BCS model,
described by the Hamiltonian,
\begin{equation}
H= \sum_{\bm p,\sigma}\xi_{\bm p}c_{\bm p\sigma}^\dagger c_{\bm p\sigma}
-U\sum_{\bm q}\sum_{\bm p,\bm p^\prime}c_{\bm p+\bm q/2\uparrow}^\dagger
c_{-\bm p+\bm q/2\downarrow}^\dagger c_{-\bm p^\prime+\bm
q/2\downarrow}c_{\bm p^\prime+\bm q/2\uparrow}.
\label{eq.1}
\end{equation}
Here, $c_{\bm p\sigma}$ is the annihilation operator of a Fermi atom
with the pseudospin $\sigma$ and the kinetic energy $\xi_{\bm p}=\varepsilon_{\bm
p}-\mu=p^2/2m-\mu$, measured from the chemical potential $\mu$ (where
$m$ is an atomic mass). $-U$ ($<0$) is an assumed tunable pairing
interaction associated with a Feshbach resonance. It is related to
the $s$-wave scattering length $a_s$ as\cite{Randeria2}
\begin{equation}
\frac{4\pi a_s}{m}=-\frac{U}{1-U\sum^{\omega_c}_{\bm p}\frac{1}{2\varepsilon_{\bm p}}},
\label{asU}
\end{equation}
where $\omega_c$ is a high-energy cutoff. Since the strength of an
interaction is usually measured in term of the scattering length $a_s$ in cold
atom physics, Eq.~(\ref{asU}) is useful in comparing theoretical results
with experiments. In this scale, the weak-coupling BCS limit and
strong-coupling BEC limit are characterized as $(k_{\rm
F}a_s)^{-1}\ll -1$ and $(k_{\rm F}a_s)^{-1}\gg +1$, respectively (where
$k_{\rm F}$ is the Fermi momentum). The region $-1\lesssim (k_{\rm
F}a_s)^{-1}\lesssim +1$ is referred to as the crossover region. The center of the crossover
region ($(k_{\rm F}a_s)^{-1}=0$) is called the unitarity limit\cite{Ho}.
\par
\begin{figure}
\centerline{\includegraphics[width=\linewidth]{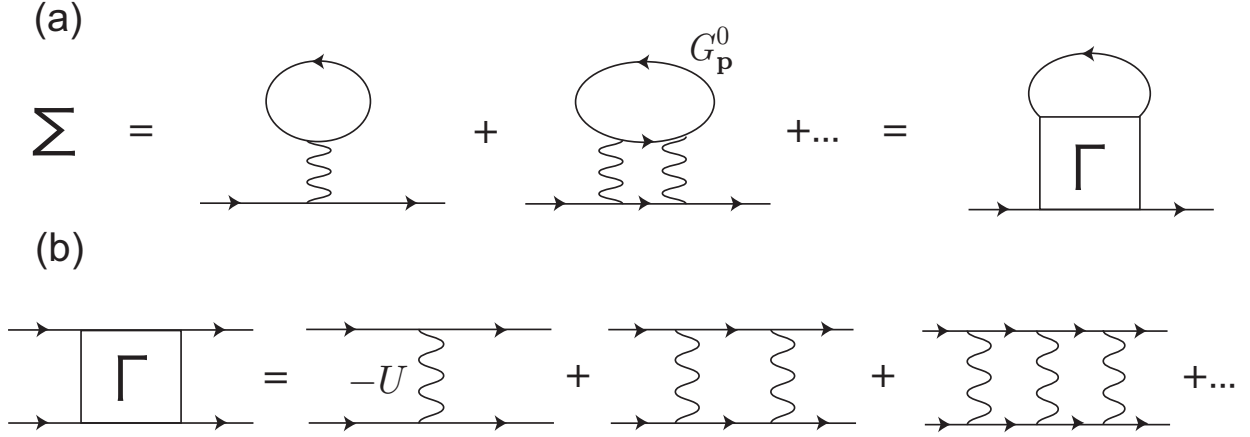}}
\caption{(a) Self-energy correction $\Sigma({\bm p},i\omega_n)$, and (b)
 particle-particle scattering matrix $\Gamma({\bm q},i\nu_n)$, in the
 $T$-matrix approximation. The solid and wavy lines represent the
 non-interacting Fermi Green's function $G^0_{\bm p}(i\omega_n)$ and
 pairing interaction $-U$, respectively.}
\label{diagram}
\end{figure}

To discuss strong-coupling effects in the BCS-BEC crossover regime above
$T_{\rm c}$, we include pairing fluctuations within the $T$-matrix
approximation\cite{Rohe,Perali}. Namely, we consider the single-particle
thermal Green's function,
\begin{eqnarray}
G_{\bm p}(i\omega_n)=\frac{1}{i\omega_n-\xi_{\bm p}-\Sigma(\bm p,i\omega_n)},
\label{Green}
\end{eqnarray}
where $\omega_n$ is the fermion Matsubara frequency. The self-energy
correction $\Sigma({\bm p},i\omega_n)$ describes effects of pairing
fluctuations, which is diagrammatically given by Fig.~\ref{diagram}(a). 
In Fig.~\ref{diagram}, the solid lines are the free fermion propagator,
\begin{equation}
G^0_{\bm p}(i\omega_n)=\frac{1}{i\omega_n-\xi_{\bm p}}.
\label{eq.free}
\end{equation}
Although this $T$-matrix theory does not treat the single-particle
Green's function self-consistently, Ref.~\cite{Perali} has shown that it
can correctly describe the smooth crossover from the BCS regime to the
BEC regime. We briefly note that the self-consistent $T$-matrix
approximation (where the full Green's function $G$ is used in stead of
$G^0$ in evaluating the self-energy) has been recently employed to study the
spectral weight and rf-spectrum in the crossover
region\cite{Haussmann}.

Summing up the diagrams in Fig.~\ref{diagram}(a), we obtain
\begin{eqnarray}
\Sigma(\bm p,i\omega_n)=T\sum_{\bm q,\nu_n}\Gamma(\bm q,i\nu_n)G_{\bm
 q-\bm p}^0(i\nu_n-i\omega_n)e^{i(\nu_n-\omega_n)\delta},
\label{selfe}
\end{eqnarray}
where $\nu_n$ is the boson Matsubara frequency. The particle-particle
scattering matrix $\Gamma({\bm q},i\nu_n)$, which describes fluctuations in the Cooper channel, is diagrammatically given by Fig.~\ref{diagram}(b). The expression is given by
\begin{eqnarray}
\Gamma({\bm q},i\nu_n)
&=&\frac{-U}{1-U\Pi(\bm q,i\nu_n)}
\nonumber
\\
&=&
{4\pi a_s \over m}
{1 \over
\displaystyle
1+{4\pi a_s \over m}
\Bigl[
\Pi(\bm q,i\nu_n)-\sum_{\bm p}{1 \over 2\varepsilon_{\bm p}}
\Bigr]
}.
\label{Gamma}
\end{eqnarray}
In the last expression, the ultraviolet divergence coming
from the contact pairing interaction has been absorbed into
the scattering length $a_s$\cite{Randeria2}. $\Pi(\bm q,i\nu_n)$ is the
pair-propagator, given by
\begin{eqnarray}
\Pi({\bm q},i\nu_n)&=&T\sum_{\bm p,\omega_n}G^0_{{\bm p}+{\bm
 q}/2}(i\nu_n+i\omega_n)G^0_{{-\bm p}+{\bm q/2}}(-i\omega_n)
\nonumber\\
&=&\sum_{\bm p}\frac{1-f(\xi_{\bm p+\bm q/2})-f(\xi_{\bm p-\bm q/2})}{\xi_{\bm p+\bm q/2}+\xi_{\bm p-\bm q/2}-i\nu_n},
\end{eqnarray}
where $f(\varepsilon)$ is the Fermi distribution function.
\par
\begin{figure}
\centerline{\includegraphics[width=\linewidth]{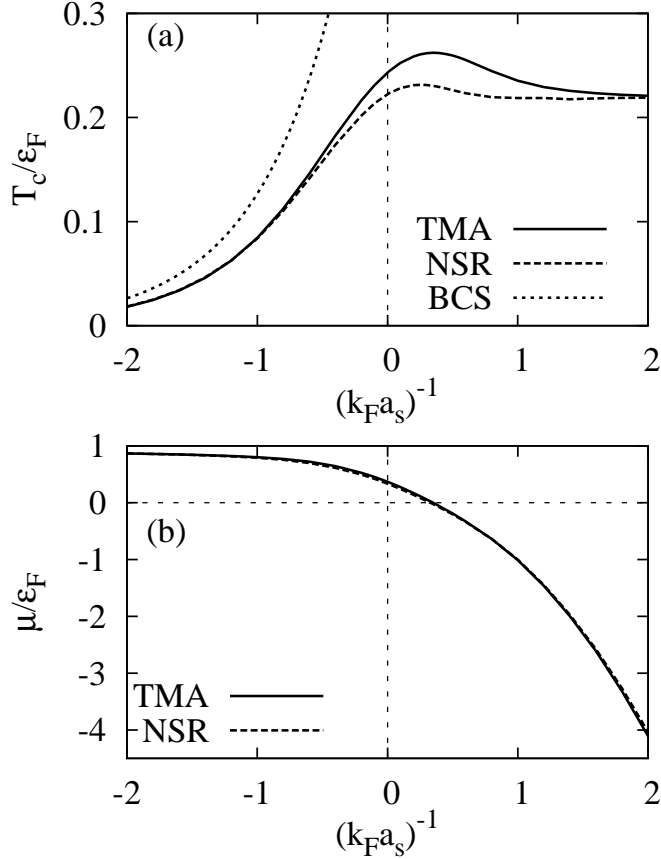}}
\caption{Self-consistent solutions of the coupled equations
 (\ref{Thouless}) and (\ref{number}) in the BCS-BEC crossover (`TMA' in
 the figure). (a) phase transition temperature $T_{\rm c}$. (b) chemical
 potential $\mu(T=T_{\rm c}$). In panel (b), $\mu$ is negative when
 $(k_{\rm F}a_s)^{-1}\ge 0.35$. `BCS' and `NSR' are the weak-coupling BCS
 result and the NSR result, respectively.}
\label{Tcmuc}
\end{figure}

To examine the pseudogap region, one needs to determine $T_{\rm
c}$\cite{Nozieres,SadeMelo,Ohashi,Perali}. The equation for $T_{\rm c}$
is obtained from the Thouless criterion\cite{Thouless}, $\Gamma({\bm
q}=0,i\nu_n=0,T=T_{\rm c})^{-1}=0$, which gives
\begin{equation}
1=-\frac{4\pi a_s}{m}\sum_{\bm p}
\left[
\frac{1}{2(\varepsilon_{\bm p}-\mu)}\tanh{\xi_{\bm p} \over 2T}-\frac{1}{2\varepsilon_{\bm p}}
\right].
\label{Thouless}
\end{equation}
As pointed out by Nozi\`eres and Schmitt-Rink\cite{Nozieres}, the
chemical potential $\mu$ deviates from the Fermi energy
$\varepsilon_{\rm F}$ in the BCS-BEC crossover. This strong-coupling
effect can be conveniently included by solving Eq.~(\ref{Thouless}),
together with the equation for the number $N$ of Fermi atoms, 
\begin{equation}
N=2T\sum_{\bm p,\omega_n}e^{i\omega_n\delta}G_{\bm p}(i\omega_n).
\label{number}
\end{equation}
We show the self-consistent solutions of the coupled equations
(\ref{Thouless}) and (\ref{number}) in Fig.~\ref{Tcmuc}.
\par

\begin{figure}
\centerline{\includegraphics[width=\linewidth]{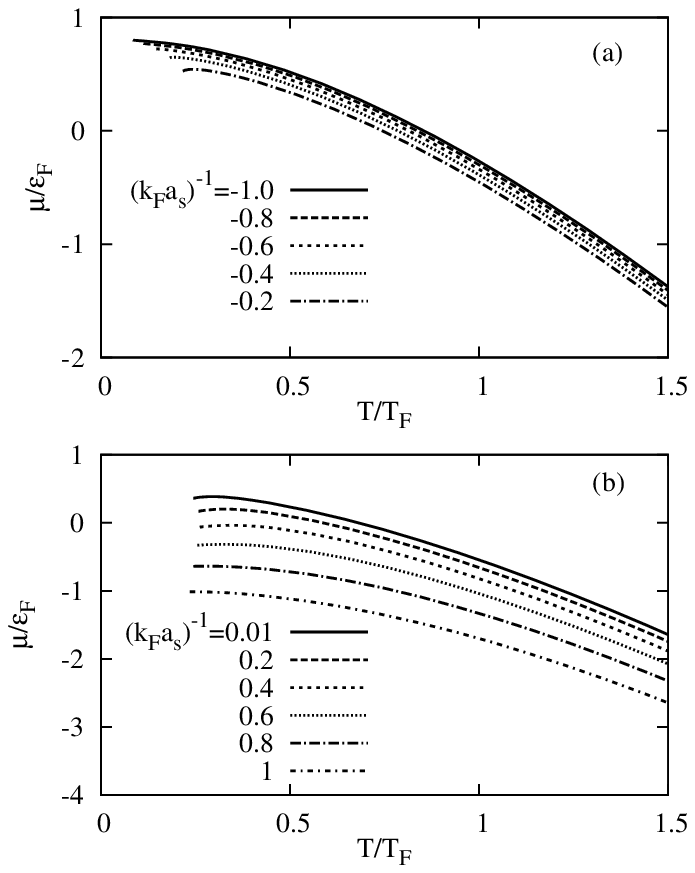}}
\caption{Calculated chemical potential $\mu$ above $T_{\rm c}$ in the BCS side
 (a) and BEC side (b). Each line starts from $T_{\rm c}$. We will use these
 results in calculating the density of states and spectral weight in
 Secs. III and IV.}
\label{muT}
\end{figure}

In the normal phase above $T_{\rm c}$, we only solve the number equation
(\ref{number}) to determine the temperature dependence of $\mu(T> T_{\rm
c})$. 
The resulting $\mu(T)$ in Fig.~\ref{muT} is used to calculate DOS $\rho(\omega)$, as well as the
spectral weight $A({\bm p},\omega)$. They are obtained from the analytic continued Green's function as, respectively,
\begin{eqnarray}
\rho(\omega)=-{1 \over \pi}\sum_{\bm p}{\rm Im}
[G(\bm p,i\omega\to \omega+i\delta)],
\label{DOS}
\end{eqnarray}
\begin{eqnarray}
A({\bm p},\omega)=-{1 \over \pi}
{\rm Im}[G(\bm p,i\omega\to \omega+i\delta)].
\label{SW}
\end{eqnarray}
The analytic continued self-energy in $G({\bm p},i\omega_n\to\omega+i\delta)$ has the form,
\begin{equation}
\Sigma(\bm p,\omega+i\delta)=\Sigma_{\rm H}+
\frac{1}{\pi}\sum_{\bm
 q}\int_{-\infty}^{\infty}dz\ \frac{n_B(z)+f(\xi_{\bm q-\bm
 p})}{z-(\omega+i\delta)-\xi_{\bm q-\bm p}}
{\rm Im}[\Gamma(\bm q,i\nu_n\to z+i\delta)],
\label{selfe2}
\end{equation}
where $n_B(\varepsilon)$ is the Bose distribution function. $\Sigma_{\rm
H}=-(U/2)\sum_{\bm p}f(\xi_{\bm p})$ is the Hartree term, and the last term
in Eq. (\ref{selfe2}) describes fluctuation correction to
single-particle excitations.
\par
Before ending this section, we comment on the $T$-matrix theory used in
this paper. In the BCS-BEC crossover literature, the so-called Gaussian
fluctuation theory developed by Nozi\`eres and Schmitt-Rink
(NSR)\cite{Nozieres,SadeMelo} has been also used. The present $T$-matrix
theory is a natural extension of this to include higher order pairing
fluctuations. Indeed, the $T_{\rm c}$-equation (\ref{Thouless}) is
common to the two theories, and the NSR number equation is also obtained
from Eq. (\ref{number}), by expanding $G_{\bm p}(i\omega_n)$ in Eq.~(\ref{number}) up to
$O(\Sigma)$, as 
\begin{equation}
G_{\bm p}^{\rm NSR}(i\omega_n)=G^0_{\bm p}(i\omega_n)+G^0_{\bm p}(i\omega_n)\Sigma(\bm p,i\omega_n)G^0_{\bm p}(i\omega_n).
\label{NSRG}
\end{equation}
The two theories essentially give the same BCS-BEC
crossover behaviors of $T_{\rm c}$ and $\mu(T=T_{\rm c})$, as shown in
Fig.~\ref{Tcmuc}. In particular, both theories correctly describe the
strong-coupling BEC limit, where the superfluid phase transition is
dominated by BEC of $N/2$ tightly bound molecules (which leads to
$T_{\rm c}=0.218T_{\rm F}$\cite{Nozieres}) and $2|\mu|$ equals the
binding energy of a two-body bound state $E_{\rm
bind}=1/ma_s^2$\cite{Leggett}. However, when one uses $G_{\bm p}^{\rm
NSR}(i\omega_n\to\omega+i\delta)$ in calculating Eq.~(\ref{DOS}), unphysical results are obtained. The NSR theory overestimates the suppression of DOS around
$\omega=0$, leading to a negative DOS around $\omega=0$
in the crossover region\cite{Tsuchiya}. 
The NSR theory also gives an unphysical divergence of DOS at
$\omega=\mu$ (although we do not explicitly show this in this
paper)\cite{Tsuchiya}. Thus, although the NSR theory can describe the
BCS-BEC crossover behaviors of $T_{\rm c}$ and $\mu$, one needs to be
careful in considering single-particle properties in the BCS-BEC
crossover. Since this problem is absent in the present $T$-matrix
theory, we employ this framework to examine DOS and the spectral weight
in this paper.

\section{Pseudogap in single-particle density of states}
\label{results}

\begin{figure}
\centerline{\includegraphics[width=\linewidth]{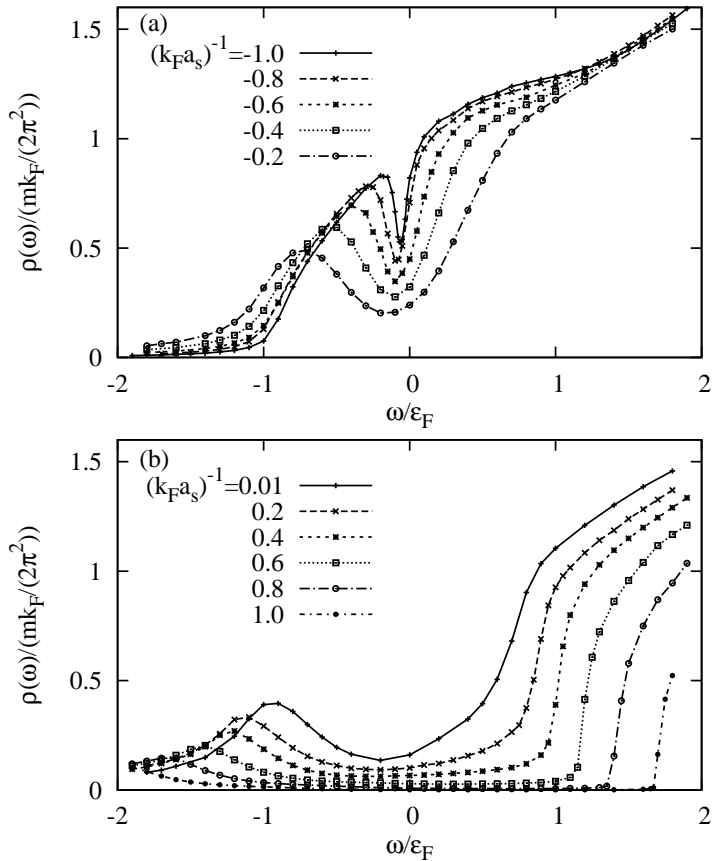}}
\caption{Density of states at $T_{\rm c}$. (a) BCS side
 ($(k_{\rm F}a_s)^{-1}<0$). (b) BEC side ($(k_{\rm F}a_s)^{-1}>0$).}
\label{dosTc}
\end{figure}

In this section, we discuss the pseudogap phenomenon in DOS.
Figure~\ref{dosTc} shows DOS in the BCS-BEC crossover at $T_{\rm
c}$. Starting from the weak-coupling BCS regime, a pseudogap
develops around $\omega=0$, as one increases the strength of the pairing
interaction. Since the superfluid order parameter vanishes at $T_{\rm
c}$, this dip structure purely originates from pairing fluctuations.
\par
The reason why the fluctuation correction described by the self-energy
in Eq.~(\ref{Green}) causes the pseudogap in DOS can be easily understood
by noting similarity between Eq.~(\ref{Green}) and the Green's
function in the mean-field BCS theory\cite{Mahan},
\begin{eqnarray}
G_{\bm p}^{\rm BCS}(i\omega_n)=-{i\omega_n+\xi_{\bm p} \over \omega_n^2+\xi_{\bm p}^2+\Delta^2},
\label{BCS}
\end{eqnarray}
where $\Delta$ is the superfluid order parameter. Assuming that pairing
fluctuations are strong around ${\bm q}=\nu_n=0$ (Note that $\Gamma({\bm
q}=0,\nu_n=0)$ diverges at $T_{\rm c}$.), we may approximate Eq.~(\ref{selfe}) to
\begin{eqnarray}
\Sigma(\bm p,i\omega_n)\simeq \Sigma_{\rm H}
-G_{-{\bm p}}^0(-i\omega_n)\Delta^2_{\rm pg},
\label{selfe3}
\end{eqnarray}
where $\Delta^2_{\rm pg}\equiv -T\sum_{\bm q,\nu_n}[\Gamma(\bm
q,i\nu_n)+U]$. 
Although $G^0_{-{\bm p}}$ in Eq.~(\ref{selfe3}) does not involve the
Hartree term $\Sigma_{\rm H}$ in the present $T$-matrix approximation, a
better approximation would involve it in evaluating $\Sigma$. In this
case, substituting Eq.~(\ref{selfe3}) into Eq.~(\ref{Green}), we obtain
\begin{eqnarray}
G_{\bm p}(i\omega_n)
=
{1 \over i\omega_n-\xi_{\bm p}+\Delta_{\rm pg}^2G_{-{\bm p}}^0(-i\omega_n)}
=
-{i\omega_n+\xi_{\bm p} \over \omega_n^2+\xi_{\bm p}^2+\Delta_{\rm pg}^2},
\label{BCS2}
\end{eqnarray}
where $\mu$ in $\xi_{\bm p}$ is replaced by $\mu+\Sigma_{\rm H}$. Since
$G^0_{-{\bm p}}(-i\omega_p)$ may be regarded as the hole Green's
function, Eq.~(\ref{BCS2}) means that pairing fluctuations induce a
particle-hole coupling. 
Comparing Eq.~(\ref{BCS2}) with Eq.~(\ref{BCS}), we find that
$\Delta_{\rm pg}$ (which describes effects of pairing fluctuations)
plays the same role as the BCS gap parameter $\Delta$. 
Actually, dynamical effects of
pairing fluctuations with ${\bm q}\ne 0$ and $\nu_n\ne 0$ smear
the clear gap structure and coherence peak known in the mean-field BCS theory. However, in Fig.~\ref{dosTc}(a), one can still
see broad peaks around $\omega/\varepsilon_{\rm F}\simeq \pm 0.2$ (which
correspond to the diverging coherence peaks at $\omega=\pm\Delta$ in the BCS theory) when
$(k_{\rm F}a_s)^{-1}\lesssim -0.4$. Although the above discussion simplifies
the treatment of pairing fluctuations, it would be helpful in
understanding the reason why pairing fluctuations give the pseudogap
structure above $T_{\rm c}$.
\par
While the pseudogapped DOS is very remarkable in the unitarity limit, it
continuously changes into a {\it fully} gapped one in the
strong-coupling BEC regime, as shown in Fig.~\ref{dosTc}(b). In the BEC
regime where $\mu$ is negative ($(k_{\rm F}a_s)^{-1}>0.35$), when we
only retain the negative $\mu$ and ignore other strong-coupling effects, the
DOS has a finite energy gap $|\mu|$ as
\begin{eqnarray}
\rho(\omega)=
\left\{
\begin{array}{ll}
0& (\omega<|\mu|),\\
{m^{3/2} \over \sqrt{2}\pi^2}
\sqrt{\omega-|\mu|}&(\omega\ge|\mu|).
\end{array}
\right.
\label{eq.becdos}
\end{eqnarray}
In the BEC limit, $2|\mu|$ equals the binding energy $E_{\rm
bind}=1/ma_s^2$ of a two-body bound state, which means that the energy gap in
Eq.~(\ref{eq.becdos}) is directly related to the molecular dissociation
energy. Since the intensity of DOS is almost absent below
$\omega/\varepsilon_{\rm F}\sim 1.4$ when $(k_{\rm F}a_s)^{-1}=+0.8$ in
Fig.~\ref{dosTc}(b), the region of $(k_{\rm F}a_s)^{-1}\gesim 0.8$ is considered to be close to 
an $N/2$ molecular gas, rather than an $N$ atomic Fermi gas.
\par
However, we note that $\rho(\omega<0)$ still has small but
{\it finite} intensity even when $(k_{\rm F}a_s)^{-1}\gesim 1.0$, as shown in Fig.~\ref{dosTc}(b), which means the existence of hole-type excitations. The finite DOS in the
negative energy region is absent when we ignore all fluctuation effects
except for the negative $\mu$ (See Eq.~(\ref{eq.becdos}).). Since the concept of hole is characteristic of many-fermion system, one finds
that, although the BEC region around $(k_{\rm F}a_s)^{-1}\simeq +1$ is
dominated by two-body molecular {\it bosons}, the character of
many-fermion system still remains to some extent there, leading to the
finite $\rho(\omega<0)$. We also find this by simply employing
Eq.~(\ref{BCS2}) to calculate DOS in the BEC regime ($\mu<0$), which
gives
\begin{eqnarray}
\rho(\omega)=
\left\{
\begin{array}{ll}
\displaystyle
{m^{3/2} \over 2\sqrt{2}\pi^2}
{\omega \over \sqrt{\omega^2-\Delta_{\rm pg}^2}}
\left[
1+{\sqrt{\omega^2-\Delta_{\rm pg}^2} \over \omega}
\right]
\sqrt{\sqrt{\omega^2-\Delta_{\rm pg}^2}-|\mu|}
& 
~~~(\omega\ge\sqrt{\Delta_{\rm pg}^2+|\mu|^2}),
\\
\displaystyle
{m^{3/2} \over 2\sqrt{2}\pi^2}
{|\omega| \over \sqrt{\omega^2-\Delta_{\rm pg}^2}}
\left[
1-{\sqrt{\omega^2-\Delta_{\rm pg}^2} \over |\omega|}
\right]
\sqrt{\sqrt{\omega^2-\Delta_{\rm pg}^2}-|\mu|}
& 
~~~(\omega\le-\sqrt{\Delta_{\rm pg}^2+|\mu|^2}).
\end{array}
\right.
\nonumber
\\
\label{DOS2}
\end{eqnarray}
When the two-body binding energy $E_{\rm bind}=1/ma_s^2~(\simeq 2|\mu|)$
is much larger than the `characteristic energy' $\Delta_{\rm pg}$, one
may ignore $\Delta_{\rm pg}$ in Eq.~(\ref{DOS2}). In this extreme BEC
limit, the upper branch in Eq.~(\ref{DOS2}) reduces to
Eq.~(\ref{eq.becdos}), and the lower one vanishes, as expected.

\begin{figure}
\centerline{\includegraphics[width=\linewidth]{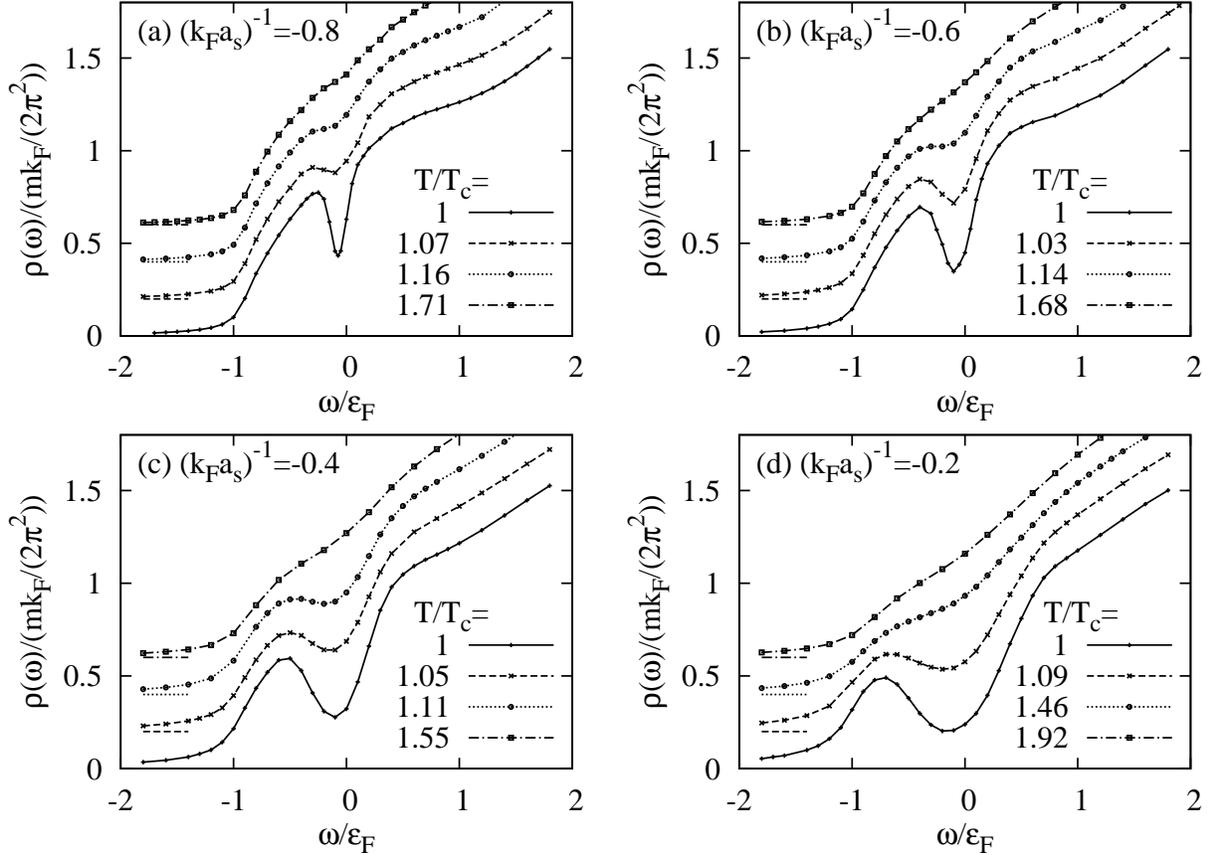}}
\caption{Temperature dependence of the density of states $\rho(\omega)$
 in the BCS side. $T_{\rm c}$ in each panel equals (a)
 0.112$\varepsilon_{\rm F}$, (b) 0.146$\varepsilon_{\rm F}$, (c)
 0.183$\varepsilon_{\rm F}$, and (d) 0.217$\varepsilon_{\rm F}$. In this
 figure and Fig.\ref{dosBEC}, we have offset the results for $T>T_{\rm
 c}$. The short horizontal line near each result is at
 $\rho(\omega)=0$.} 
\label{dosBCS}
\end{figure}

\begin{figure}
\centerline{\includegraphics[width=\linewidth]{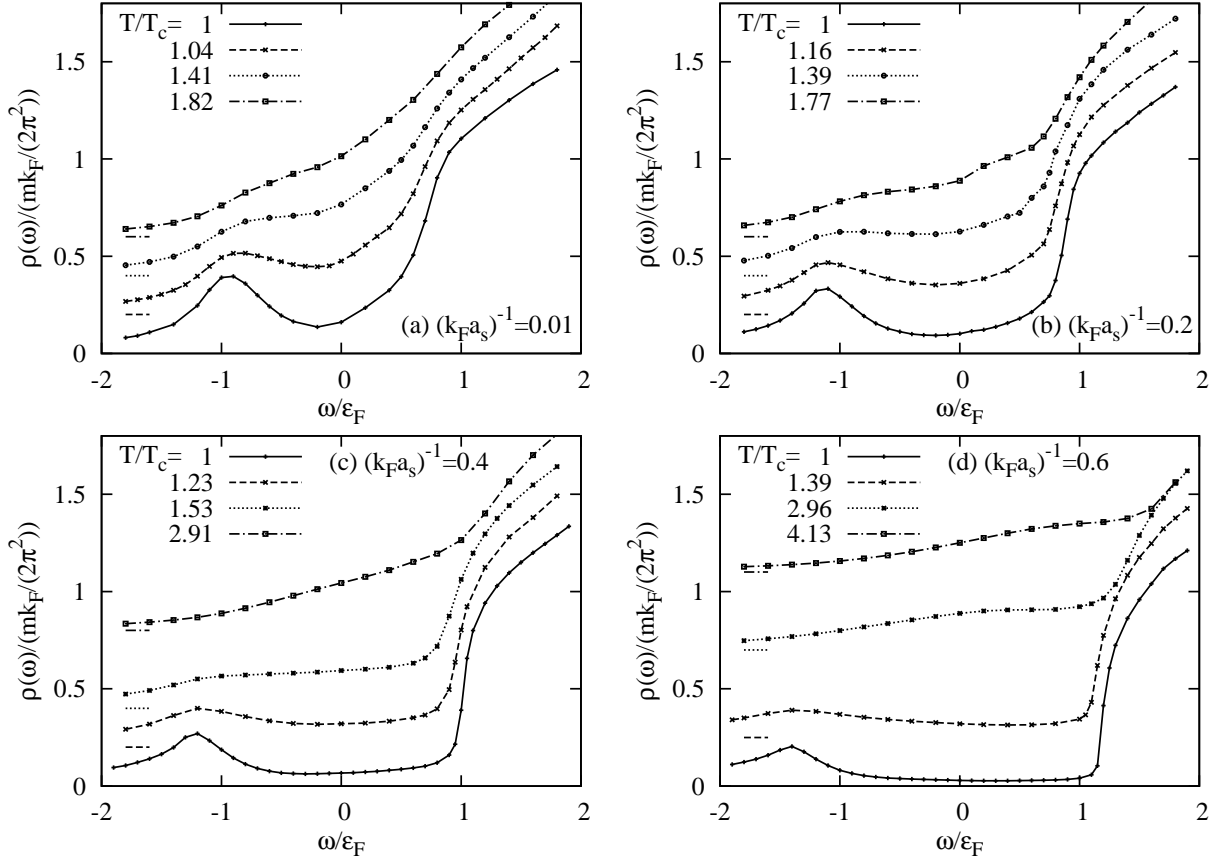}}
\caption{Temperature dependence of DOS in the BEC side.
$T_{\rm c}$ in each panel equals (a) 0.244$\varepsilon_{\rm F}$, (b)
 0.259$\varepsilon_{\rm F}$, (c) 0.262$\varepsilon_{\rm F}$, and (d)
 0.255$\varepsilon_{\rm F}$.
}
\label{dosBEC}
\end{figure}

\begin{figure}
\centerline{\includegraphics[width=0.8\linewidth]{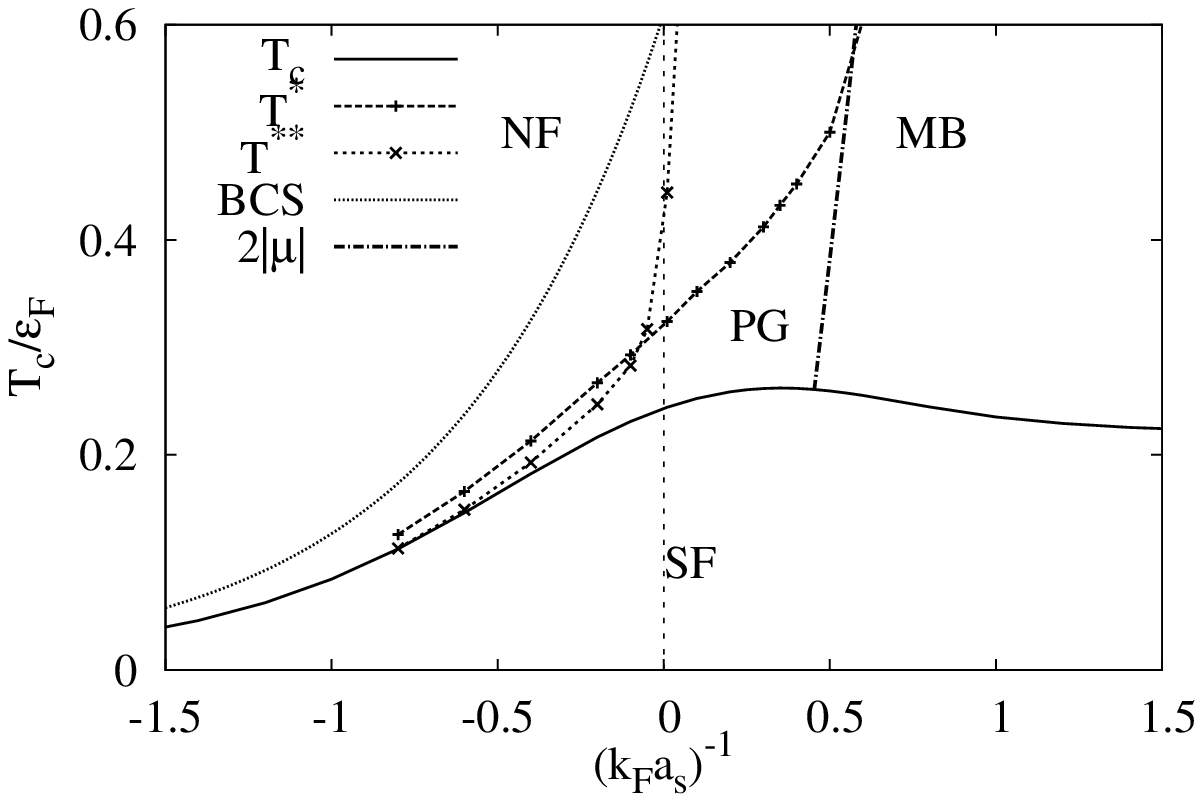}}
\caption{Pseudogap temperature $T^*$ determined from DOS in the BCS-BEC
 crossover. We also plot another pseudogap temperature $T^{**}$ where
 the double peak structure in the spectral weight vanishes. `BCS' is
 $T_{\rm c}=(8\gamma/\pi e^2)\varepsilon_{\rm F} e^{\pi/2k_{\rm F}a_s}$
 in the mean-field BCS theory (where $\gamma=1.78$)\cite{Pethick}. $T^*$
 or $T^{**}$ gives the boundary between the pseudogap regime (PG) and
 normal Fermi gas regime (NF). $2|\mu|$ $(\simeq E_{\rm bind})$ in the
 BEC regime gives the characteristic temperature below which thermal
 dissociation of bound molecules are suppressed. Namely, $T\simeq
 2|\mu|$ physically describes the boundary between PG and molecular Bose
 gas regime (MB).}
\label{phdgm}
\end{figure}
\par
Figures~\ref{dosBCS} and \ref{dosBEC} show DOS above $T_{\rm c}$. The
pseudogap structure in DOS becomes obscure at high temperatures due to
weak pairing fluctuations. The dip structure eventually vanishes at a
certain temperature, which we define as the pseudogap temperature $T^*$\cite{noteTT}.
\par
Figure~\ref{phdgm} shows the resulting pseudogap temperature $T^*$ in
the BCS-BEC crossover. Starting from the weak-coupling BCS regime, $T^*$
monotonically increases. However, $T^*$ is still lower than $T_{\rm c}$
calculated in the mean-field BCS theory (`BCS' in
Fig.~\ref{phdgm}). Although the mean-field $T_{\rm c}$ is sometimes
considered as a characteristic temperature where preformed pairs are
formed, our result shows that the pseudogap actually starts to develop
in DOS from lower temperature.
\par
We note that, although the fact that the pseudogap disappears at $T^*$
is common to the entire BCS-BEC crossover region, the detailed way of
disappearance is somehow different in between the BCS regime and
crossover-BEC regime. In Fig.~\ref{dosBCS}(a), the pseudogap around
$\omega=0$ is simply filled up at high temperatures. The shape of DOS
then becomes close to DOS of a free Fermi gas,
\begin{eqnarray}
\rho(\omega)={m^{3/2} \over \sqrt{2}\pi^2}
\sqrt{\omega+\mu}~~~~~(\omega\ge-\mu).
\label{eq.becdos2}
\end{eqnarray}
Namely, as far as we consider DOS, the system may be regarded as a
(weakly interacting) normal Fermi gas above $T^*$. On the other
hand, in the BEC side shown in Fig.~\ref{dosBEC}, in addition to the
enhancement of DOS around $\omega=0$, the lower peak is suppressed at
high temperatures. In the unitarity limit (Fig.~\ref{dosBEC}(a)), when
the pseudogap is completely filled up, DOS still has a different
shape from DOS of a free Fermi gas. In the BEC regime where $\mu<0$, 
Figs.~\ref{dosBEC}(c) and (d) show that DOS above $T^*$ has a
finite intensity in the negative energy region, in contrast to
Eq.~(\ref{eq.becdos}). These results indicate that pairing fluctuations
still affect single-particle excitations above $T^*$ in the BEC side,
although the depression of DOS around $\omega=0$ is absent. Indeed, in
Sec.~\ref{spweight}, we will show an evidence of such fluctuation effects
in the spectral weight in this regime.

\begin{figure}
\centerline{\includegraphics[width=\linewidth]{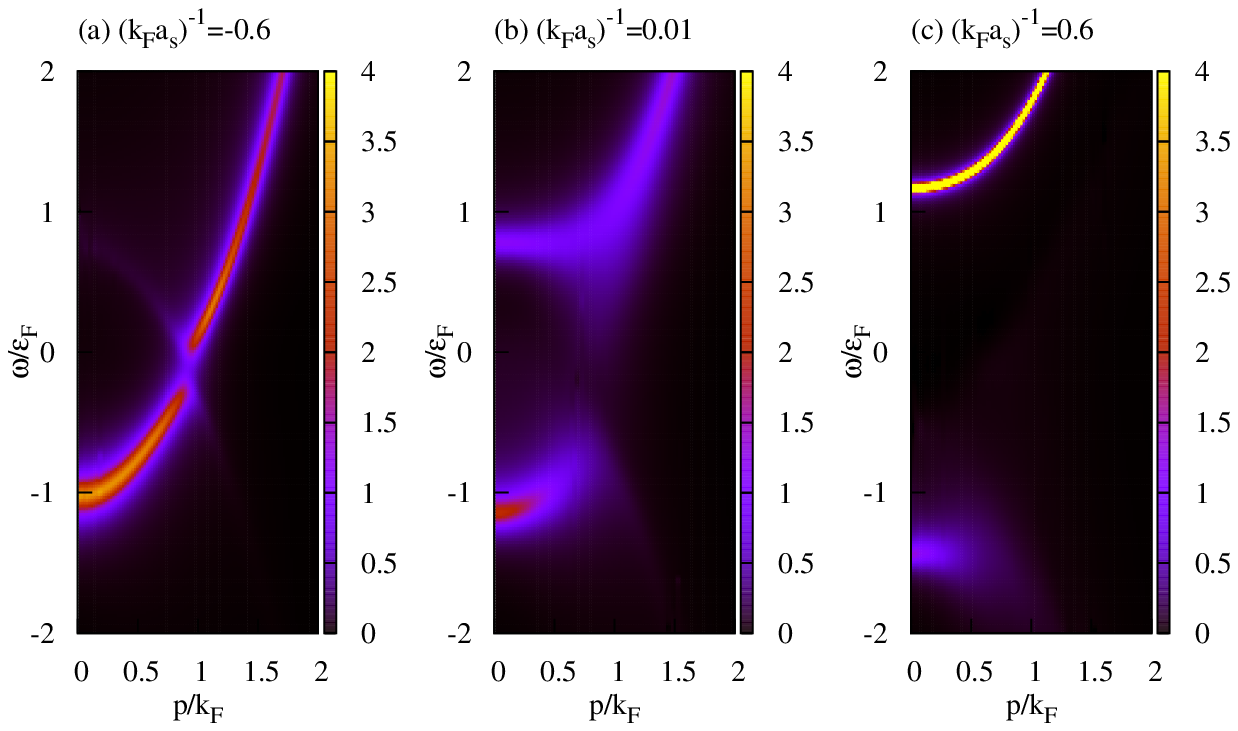}}
\caption{Calculated intensity of the spectral weight $A(\bm p,\omega)$ at $T_{\rm c}$ in the energy-momentum plane.  (a) BCS side ($(k_{\rm
 F}a_s)^{-1}=-0.6$). (b) Unitarity limit ($(k_{\rm F}a_s)^{-1}=0.01$). (c) BEC
 side ($(k_{\rm F}a_s)^{-1}=0.6$).} 
\label{swTc}
\end{figure}
\begin{figure}
\centerline{\includegraphics[width=\linewidth]{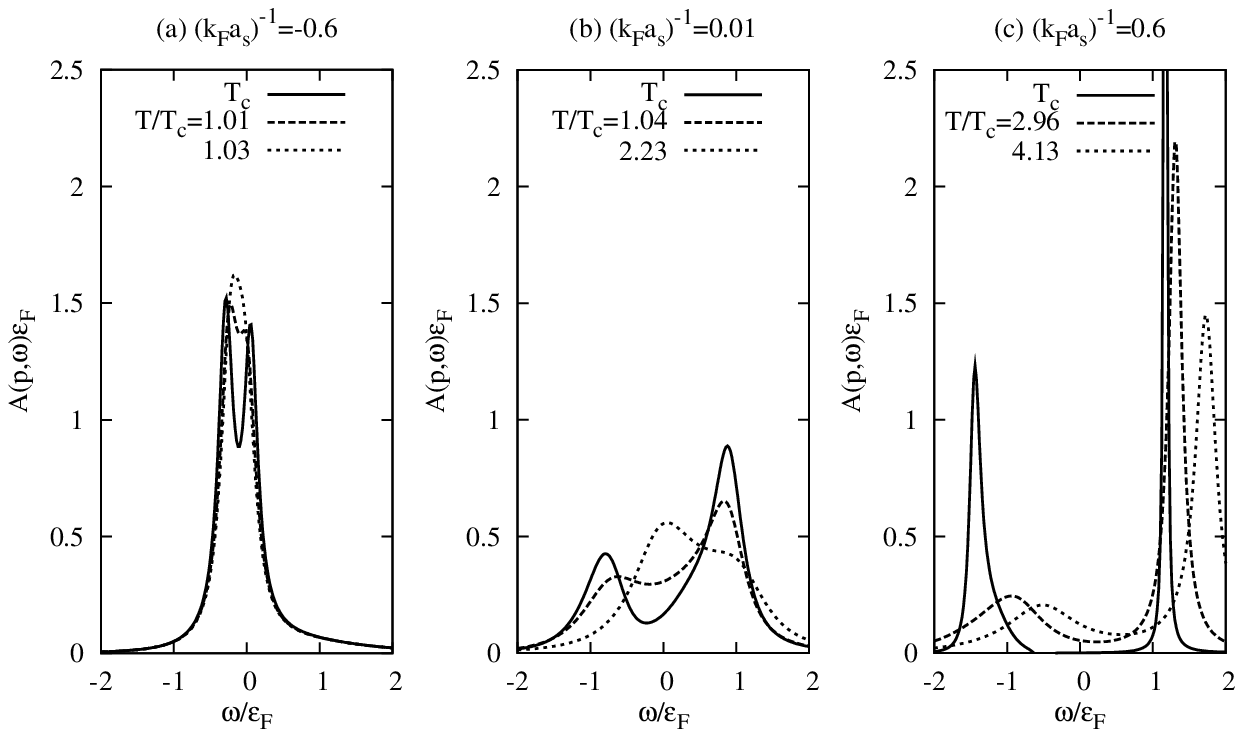}}
\caption{Spectral weight $A(\bm p,\omega)$ as a function of $\omega$. In
 each panel, we take the momentum where the peak-to-peak energy becomes
 minimum: (a) $p/k_{\rm F}=0.91$, (b) 0.83, and (c) 0.01.}
\label{swTc2}
\end{figure}

\section{Pseudogap in spectral weight}
\label{spweight}

It has been pointed out\cite{Janko,Rohe,Yanase,Perali} that pairing
fluctuations cause a BCS-type double peak structure in the
single-particle spectral weight $A({\bm p},\omega)$. In
this section, we examine how this strong-coupling effect is related to
the pseudogap in DOS discussed in Sec.~\ref{results}.
\par
Figure~\ref{swTc} shows the intensity of spectral weight $A({\bm
p},\omega)$ at $T_{\rm c}$ in the energy-momentum plane. In the BCS side
(panel (a)), in addition to the particle branch at $\omega\simeq\xi_{\bm
p}$, we can see a weak peak line of a hole branch at
$\omega\simeq-\xi_{\bm p}$. The intensity of the particle branch is
suppressed around $\omega=0$, where it intersects with
the hole branch and the level repulsion between them occurs. 
The resulting structure is similar to the BCS spectral
weight\cite{Janko,Rohe,Yanase,Perali,note}, given by\cite{Mahan}
\begin{equation}
A_{\rm BCS}(\bm p,\omega)=u_{\bm p}^2\delta(\omega-E_{\bm p})+v_{\bm
 p}^2\delta(\omega+E_{\bm p}),
\label{sfBCS}
\end{equation}
where $u_{\bm p}^2=(1+\xi_{\bm p}/E_{\bm p})/2$, $v_{\bm
p}^2=(1-\xi_{\bm p}/E_{\bm p})/2$, and $E_{\bm p}=\sqrt{\xi_{\bm
p}^2+\Delta^2}$ is the Bogoliubov quasiparticle excitation spectrum. 
For a given momentum $p$, $A_{\rm BCS}(\bm p,\omega)$ has two peaks at
$\omega=\pm E_{\bm p}$.
The negative energy branch at $\omega=-E_{\bm p}$ given by the second term
in Eq.~(\ref{sfBCS}) is dominant in the low momentum region $p\ll p_{\rm
F}$ (where $u_{\bm p}\ll v_{\bm p}$). 
On the other hand, the positive energy branch ($\omega=+E_{\bm p}$) becomes
crucial when $p\gg k_{\rm F}$ (where $u_{\bm p}\gg v_{\bm p}$). 
The existence of two branches can be understood from the Bogoliubov
transformation $c_{\bm p\uparrow}=u_{\bm p}\gamma_{\bm p\uparrow}+v_{\bm
p}\gamma_{-\bm p\downarrow}^\dagger$ ($\gamma_{\bm p\sigma}$ is an
annihilation operator of a quasiparticle with momentum $\bm p$ and spin
$\sigma$), which indicates that the annihilation of an atom is accompanied by creation and annihilation of Bogoliubov excitations\cite{Griffin}. 
The minimum energy gap $2\Delta$ between the two branches $\omega=\pm
E_{\bm p}$ is obtained at the Fermi level $p=k_{\rm F}$. 
Since the simplified Green's function in
Eq.~(\ref{BCS2}) has the same form as Eq.~(\ref{BCS}), Eq.~(\ref{BCS2})
gives rise to the spectral weight similar to the BCS type in
Eq.~(\ref{sfBCS}), where the superfluid gap $\Delta$ is now replaced by
the pseudogap $\Delta_{\rm pg}$, describing effects of pairing
fluctuation. The minimum value $2\Delta_{\rm pg}$ of the pseudogap
energy is obtained at $p\simeq k_{\rm F}$ in this case.
From this reason, the double peak
structure in Fig.~\ref{swTc}(a) is found to come from the particle-hole
coupling due to strong pairing fluctuations\cite{Perali}. In addition,
they also induce finite lifetime of quasiparticle excitations, leading
to finite widths of the two peaks in $A({\bm p},\omega)$\cite{Perali}. 
This feature is absent in the BCS spectral weight in Eq.~(\ref{sfBCS}), which has two $\delta$-functional peaks at $\omega=\pm E_{\bm p}$. 
As a result, $A({\bm p},\omega)$ at the momentum where the minimum peak-to-peak energy is obtained has finite spectral weight between the two peaks, as shown 
in Fig.~\ref{swTc2}, giving finite intensity of DOS inside the pseudogap. 
This {\it gapless} double peak structure is referred
to as the pseudogap in the spectral weight in the
literature\cite{Janko,Rohe,Yanase,Perali}.
\par
This pseudogap structure in the spectral weight becomes remarkable, as one approaches the
unitarity limit. In this limit, strong pairing fluctuations also broaden
the spectral peaks, as shown in Fig.~\ref{swTc}(b). In
the BEC regime (Fig.~\ref{swTc}(c)), the peak width of the upper branch
shrinks. This is because the BEC regime is well described by a gas of
tightly bound molecules, so that the upper branch simply describes their
dissociation. Since the molecular formation simply occurs within
two-body physics in the BEC limit, the peak of the lower branch
(which is an evidence of many-body physics) is low and broad in
Fig.~\ref{swTc}(c).
\par

These different behaviors of upper and lower peaks in the BEC regime can be directly understood from the imaginary part of the self-energy correction. Using the fact that the particle-particle scattering matrix $\Gamma$ reduces to the Bose Green's function in the BEC limit as\cite{Perali}
\begin{equation}
\Gamma(\bm q,i\nu_n)=\frac{8\pi}{m^2a_s}\frac{1}{i\nu_n-E_{\bm q}^B}
\label{bGreen}
\end{equation}
(where $E_{\bm q}^B=q^2/4m-\mu_B$ is the energy of a molecule measured from the molecular chemical potential $\mu_B\simeq 2\mu+1/(ma_s^2)\simeq 0$), we can approximately evaluate the imaginary part of the analytic continued self-energy in Eq.~(\ref{selfe2}) as
\begin{eqnarray}
{\rm Im}\Sigma(\bm p,\omega+i\delta)&=&-\frac{8\pi^2}{m^2a_s}\sum_{\bm
 q}n_B(E_{\bm q}^B)\delta\left(\omega-(E_{\bm q}^B-\xi_{\bm q-\bm p})\right),
\nonumber\\
&=&-\frac{4T}{a_s
p}\ln\left[\frac{1-\exp\left\{-\beta\left(\frac{3p^2}{2m}+\Delta\omega+\frac{2p}{\sqrt{m}}\sqrt{\frac{p^2}{2m}+\Delta\omega}-\mu_B\right)\right\}}{1-\exp\left\{-\beta\left(\frac{3p^2}{2m}+\Delta\omega-\frac{2p}{\sqrt{m}}\sqrt{\frac{p^2}{2m}+\Delta\omega}-\mu_B\right)\right\}}\right]
\nonumber
\\
&\times&\theta({p^2 \over 2m}+\omega_{\rm th}-\omega),
\label{eq22}
\end{eqnarray}
where $\Delta\omega=\omega_{\rm th}-\omega$, and $\omega_{\rm th}=\mu-\mu_B\simeq -1/2ma_s^2$. Since ${\rm Im}\Sigma(\bm p,\omega+i\delta)$ directly gives the peak width of the spectral weight, the first line in Eq.~(\ref{eq22}) indicates that, in the BEC regime, the peak widths are dominated by molecules excited thermally with finite center of mass momentum $\bm q\neq 0$. Since Eq. (\ref{eq22}) vanishes when $\omega>p^2/2m+\omega_{\rm th}\simeq p^2/2m-1/2ma_s^2$, the upper branch around $\omega=\xi_{\bm p}$ ($>0$) appears as a sharp delta-function peak in the spectral weight in the BEC limit. This is consistent with the sharp upper peak in Fig.~\ref{swTc2}(c).
\par
On the other hand, expanding Eq. (\ref{eq22}) around the lower branch, $\omega=\xi_{\bm p}$, one obtains
\begin{equation}
{\rm Im}\Sigma(\bm
 p,\omega+i\delta)\simeq\frac{4T}{a_sp}\ln\left(\frac{m}{4Tp^2}\delta\omega^2\right),
\label{selfBEC2}
\end{equation}
where $\delta\omega=\omega-(-\xi_{\bm p})$. 
Equation~(\ref{selfBEC2}) shows that the imaginary part of the
self-energy logarithmically diverges along the lower branch $\omega=-\xi_{\bm p}$. Thus, the lower peak is smeared out in the BEC limit. 


\begin{figure}
\centerline{\includegraphics[width=\linewidth]{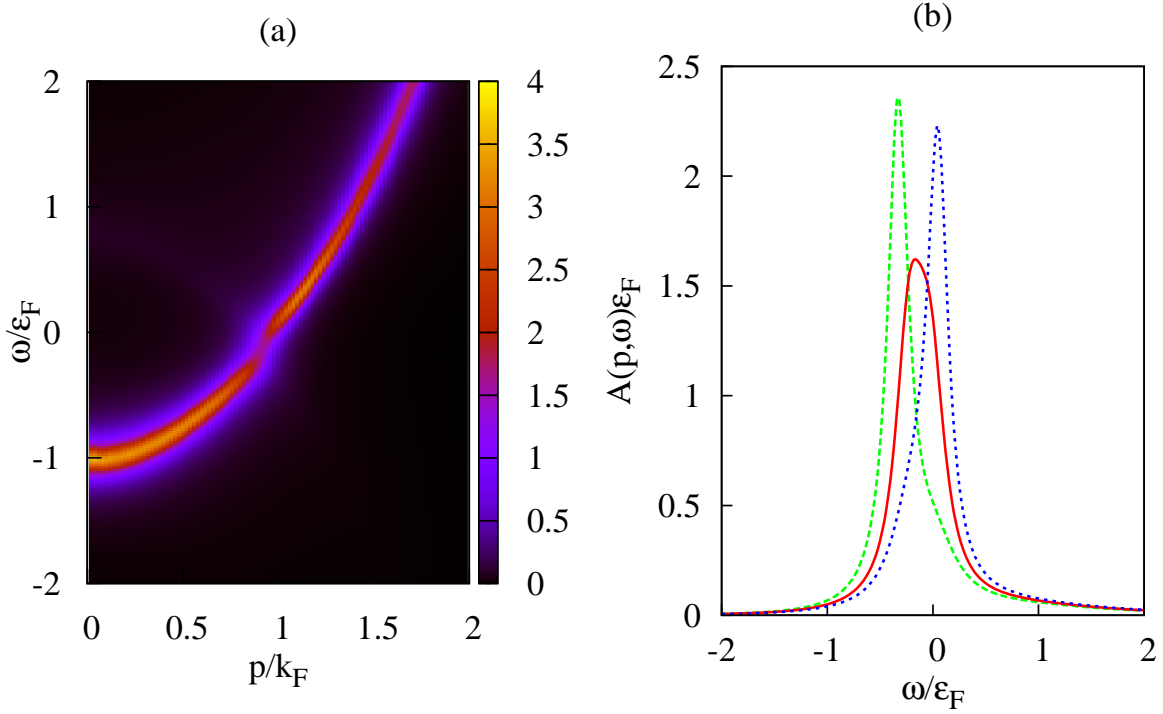}}
\caption{(a) Intensity of the spectral weight $A({\bm p},\omega)$ in the BCS
 side ($(k_{\rm F}a_s)^{-1}=-0.6$). We set $T/T_{\rm c}=1.03$, at which the dip
 structure can be clearly seen in DOS. (b) $A({\bm p},\omega)$ as a function
 of $\omega$.  The momentum $p$ is taken to be $p/k_{\rm F}=0.91$ (solid
 line), 0.83 (dashed line), and 0.97 (dotted line).}
\label{swBCS}
\end{figure}

\begin{figure}
\centerline{\includegraphics[width=\linewidth]{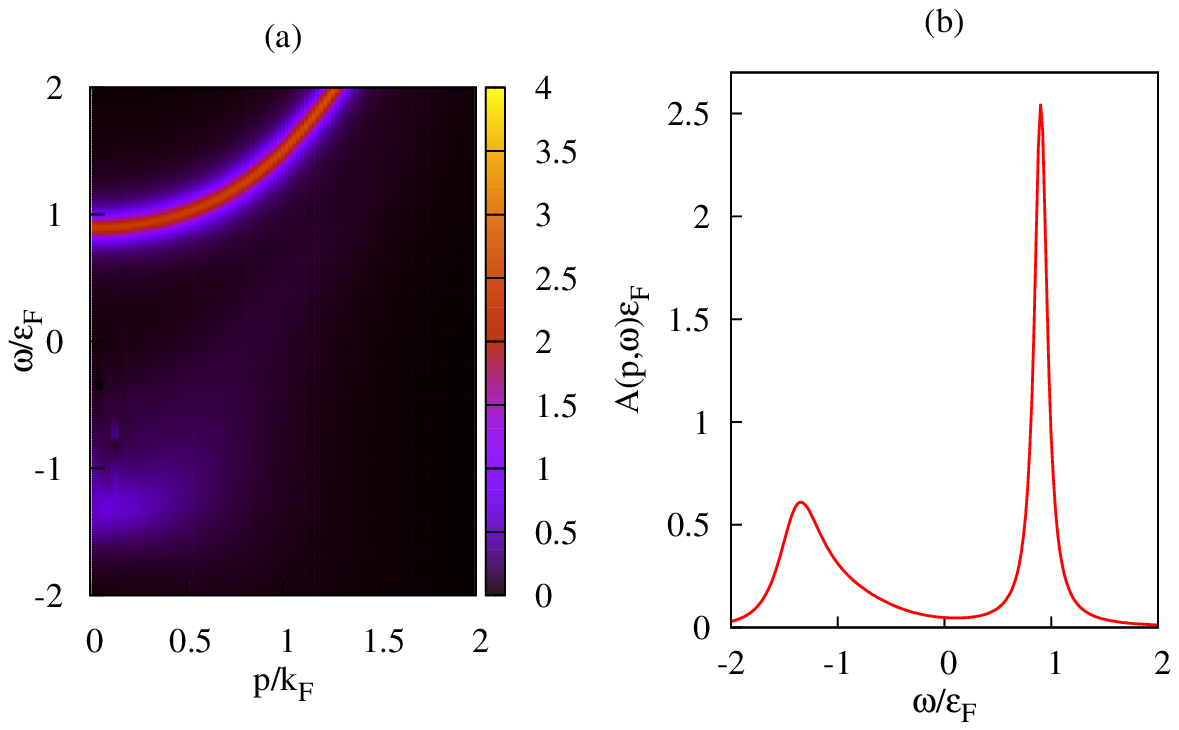}}
\caption{(a) Intensity of spectral weight $A({\bm p},\omega)$ in the BEC side
 ($(k_{\rm F}a_s)^{-1}=+0.4$). We take $T/T_{\rm c}=1.53$, at which the
 pseudogap structure is absent in DOS.  (b) $A({\bm p},\omega)$ as a function
 of $\omega$ at $p=0.01k_{\rm F}$.}
\label{swBEC}
\end{figure}

As one increases the temperature, Fig.~\ref{swTc2} shows that the double
peak structure gradually becomes obscure to eventually vanish at a
certain temperature $(\equiv T^{**})$. Regarding $T^{**}$ as another
pseudogap temperature\cite{noteTTT}, 
one might expect that it is deeply related to
$T^*$ defined from DOS, because DOS is given by the momentum summation
of the spectral weight. However, when we compare $T^{**}$ with $T^*$ in
the BCS-BEC crossover, they are very different from each other, as shown
in Fig.~\ref{phdgm}. While one sees $T^*>T^{**}$ in the BCS
side\cite{Janko}, $T^{**}$ becomes higher than $T^*$ in the BEC side
($(k_{\rm F}a_s)^{-1}\gesim -0.07$).
\par
In the BCS side, when $T\gesim T^{**}$, since pairing fluctuations are
still strong near the Fermi surface, the single peak in the spectral
weight at $p\simeq\sqrt{2m\mu}$ is broad and the peak height is low,
compared with the cases of higher and lower momenta, as shown in
Fig.~\ref{swBCS}. This low peak height at $p\simeq\sqrt{2m\mu}$ directly
affects the density of states around $\omega=0$, leading to the dip or pseudogap structure in $\rho(\omega)$ in the region $T^{**}\le T\le T^*$. We briefly note that
the result of $T^*>T^{**}$ in the BCS side agrees with the previous
work\cite{Janko}.
\par
On the other hand, although the double peak structure still exists when
$T> T^*$ in the BEC side, the intensity of the lower peak is 
very weak and broad (See
Fig.~\ref{swBEC}.), because the system is close to a gas of two-body
bound molecules. Thus, the existence of lower peak is easily smeared out
in the momentum summation in calculating DOS, $\rho(\omega)=\sum_{\bm p}A({\bm p},\omega)$.
\par
To see the physical backgrounds of $T^*$ and $T^{**}$, it is convenient
to recall that, when pairs are formed above $T_{\rm c}$, the
lifetime of Fermi excitations becomes short 
due to strong tendency to form pairs, leading to a broad
quasi-particle peak in the spectral weight $A({\bm p},\omega)$. In
addition, preformed pairs also induce the particle-hole coupling, which
gives the double peak structure in $A({\bm p},\omega)$. Between the two
effects associated with pair formation, while $T^{**}$ is directly related to the latter by
definition, the former is crucial for $T^*$: In the BCS regime, since
the peak-to-peak energy in $A({\bm p},\omega)$ is small, the double-peak
pseudogap structure is easily smeared out by the lifetime effect,
namely, the broadening of two peaks. On the other hand, DOS around
$\omega=0$ is suppressed, when the height of quasiparticle peak at
$\omega\simeq 0$ is lowered by the broadening effect. As a result, one
obtains $T^*>T^{**}$ in the BCS regime, and one may use $T^*$ as the
characteristic temperature where preformed pairs are formed. The double
peak structure can be clearly seen in $A({\bm p},\omega)$ in the
crossover-BEC regime, because the peak-to-peak energy becomes larger
than the peak widths. However, as discussed previously, the lower peak
becomes very broad and the weight becomes small in the BEC regime,
reflecting that the system is close to a gas of two-body bound
molecules. Thus, one cannot see the dip structure in DOS even when the
particle-hole coupling induce the double peak structure in $A({\bm
p},\omega)$ below $T^{**}$. As one further decreases the temperature,
the lower peak in $A({\bm p},\omega)$ shrinks and the peak height
increases, because the system approaches the superfluid phase. This
clearly enhances the intensity of DOS in the negative energy region,
leading to the dip structure below $T^* (<T^{**})$.  
\par
The different behaviors of two pseudogap temperatures
$T^*$ and $T^{**}$ imply that the pseudogap region may depend on what we
measure. When we consider a quantity where DOS is crucial, $T^*$ would
give the boundary between the pseudogap region and normal Fermi gas
regime. On the other hand, when we consider a quantity dominated by the
spectral weight, $T^{**}$ would be observed as the boundary between the
two regions. While the specific heat is an example of the former
quantity, the recent photoemission-type experiment\cite{Stewart} is
considered to be a latter example.
\par
We note that, when the temperature is lower than the binding energy
$E_{\rm bind}\simeq 2|\mu|$ of a two-body bound molecule in the BEC
regime, thermal dissociation of molecules is suppressed. In this sense,
one may regard this regime as a molecular Bose gas, rather than a
(strongly-correlated) Fermi gas. Including this, we obtain the phase
diagram in Fig.~\ref{phdgm}. In this figure, the pseudogap regime is the
region surrounded by $T^*$ or $T^{**}$, $T_{\rm c}$ and $2|\mu|$. We
briefly note that except for $T_{\rm c}$, other temperatures $T^*$,
$T^{**}$, and $T=2|\mu|$, are all crossover temperatures without
accompanied by any phase transition.

\par
\section{summary}
\par
To summarize, we have investigated the pseudogap behaviors of an
ultracold Fermi gas in the BCS-BEC crossover above $T_{\rm c}$. We have
calculated the single-particle density of states (DOS), as well as
the single-particle spectral weight, including pair fluctuations within
the framework of $T$-matrix approximation. We showed how the pseudogap
structure appears/disappears in DOS above $T_{\rm c}$ in the BCS-BEC crossover
region. Starting from the weak-coupling BCS regime, while the pseudogap in 
DOS becomes remarkable near the unitarity limit, it continuously changes into a
fully gapped DOS in the BEC regime.
\par
We determined the pseudogap temperature $T^*$ as the temperature when
the dip structure in DOS disappears. We also introduced another pseudogap
temperature $T^{**}$ at which the double peak structure in the spectral
weight vanishes. We showed that, although both the dip structures in DOS
and the double peak structure in the spectral weight originate from
pairing fluctuations, their values are very different from each other in
the BCS-BEC crossover. While one finds $T^*>T^{**}$ in the BCS
side ($(k_{\rm F}a_s)^{-1}\lesssim 0$), $T^{**}$ becomes much higher
than $T^{*}$ in the BEC side ($(k_{\rm F}a_s)^{-1}\gesim 0$). This means
that the pseudogap region may depend on the physical quantities which we
measure. In particular, since the recent photoemission-type
experiment\cite{Stewart} is related to the spectral weight, one expects
that $T^{**}$ would work as the pseudogap temperature in this
experiment. Including $T^*$ and $T^{**}$, we determined the pseudogap
region in the BCS-BEC phase diagram with respect to temperature and the
strength of pairing interaction. Since the pseudogap effects are crucial
in understanding strong-coupling Fermi superfluids, our results would be
useful in the search for the pseudogap region in the BCS-BEC crossover
regime of ultracold Fermi gases.

\acknowledgments
\par
We would like to thank A. Griffin for valuable discussions and
comments. This work was supported by a Grant-in-Aid for Scientific
research from MEXT in Japan (18043005,20500044).
\par


\begin{thebibliography}{}
\bibitem{noteTT} In this paper, we determine $T^*$ as the temperature
	when the region of $d\rho(\omega)/d\omega<0$ disappears.
\bibitem{note} When we plot $A_{\rm BCS}({\bm p},\omega)$ in the energy-momentum
	plane, the BCS gap symmetrically appears as $\omega=\pm\Delta$
	at $p=k_{\rm F}$. In Fig.~\ref{swTc}(a), although the region
	where the peak line is suppressed is not symmetric with respect
	to $\omega=0$, this is simply due to the non-interacting Green's
	function $G^0_{\bm p}(i\omega_n)$ used to calculate the self-energy. When the
	shift of the chemical potential by the self-energy correction is
	correctly included in $G^0_{\bm p}(i\omega_n)$, it has been shown that the
	pseudogapped region in $A({\bm p},\omega)$ appears symmetrically
	with respect to $\omega=0$. For more details, see
	Ref.~\cite{Perali}.
\bibitem{noteTTT} In this paper, we determine $T^{**}$ as the
	temperature when the double peak structure of $A({\bm
	p},\omega)$ at $p=\sqrt{2m\mu}$ vanishes, when $\mu>0$. In the
	BEC regime where $\mu\le 0$, we use the spectral weight at $p=0$
	for this purpose.
\end{thebibliography}
\end{document}